\documentclass[aps,prstab,reprint,groupedaddress]{revtex4-2}

\usepackage{ulem}
\usepackage{graphicx}      
\usepackage{natbib}        
\usepackage{amsmath,amssymb}
\usepackage{xcolor}
\usepackage{color}
\usepackage[ansinew]{inputenc}
\usepackage{amsfonts}
\usepackage{amsmath}
\usepackage{lipsum}

\begin{document}
\title{RF regulation with superconducting cavities and beam operation using a frequency shifted cavity}

\author{S.~Pfeiffer} 
\email{sven.pfeiffer@desy.de}
\affiliation{Deutsches Elektronen-Synchrotron DESY, Notkestr. 85, 22607 Hamburg, Germany }

\author{V.~Ayvazyan}
\affiliation{Deutsches Elektronen-Synchrotron DESY, Notkestr. 85, 22607 Hamburg, Germany }

\author{J.~Branlard}
\affiliation{Deutsches Elektronen-Synchrotron DESY, Notkestr. 85, 22607 Hamburg, Germany }

\author{T.~Buettner}
\affiliation{Deutsches Elektronen-Synchrotron DESY, Notkestr. 85, 22607 Hamburg, Germany }

\author{S.~Choroba}
\affiliation{Deutsches Elektronen-Synchrotron DESY, Notkestr. 85, 22607 Hamburg, Germany }

\author{B.~Faatz}
\affiliation{Shanghai Advanced Research Institute, Chinese Academy of Sciences, Haike Road 99,
Shanghai 201210, China}

\author{K.~Honkavaara}
\affiliation{Deutsches Elektronen-Synchrotron DESY, Notkestr. 85, 22607 Hamburg, Germany }

\author{V.~Katalev}
\affiliation{Deutsches Elektronen-Synchrotron DESY, Notkestr. 85, 22607 Hamburg, Germany }

\author{H.~Schlarb}
\affiliation{Deutsches Elektronen-Synchrotron DESY, Notkestr. 85, 22607 Hamburg, Germany }

\author{C.~Schmidt}
\affiliation{Deutsches Elektronen-Synchrotron DESY, Notkestr. 85, 22607 Hamburg, Germany }

\author{S.~Schreiber}
\affiliation{Deutsches Elektronen-Synchrotron DESY, Notkestr. 85, 22607 Hamburg, Germany }

\author{A.~Sulimov}
\affiliation{Deutsches Elektronen-Synchrotron DESY, Notkestr. 85, 22607 Hamburg, Germany }

\author{E.~Vogel}
\affiliation{Deutsches Elektronen-Synchrotron DESY, Notkestr. 85, 22607 Hamburg, Germany }

\author{H.~Weise}
\affiliation{Deutsches Elektronen-Synchrotron DESY, Notkestr. 85, 22607 Hamburg, Germany }

\date{\today}


\begin{abstract}
The free-electron laser FLASH at DESY and the European XFEL are operated with superconducting radio frequency cavities and supply beam to several user experiments. The switching time between experiments is limited to dozens of microseconds. This contribution will show a regulation with a frequency shifted superconducting cavity to manipulate and change the accelerating properties of electron bunches with 250\;kHz.The main challenge of the concept presented in this contribution can be summarized in this statement: finding a way to modulate the energy of individual bunches in a single-source multiple-cavity scheme for a potential CW upgrade of the EuXFEL. 
\end{abstract}


\maketitle


\section{Introduction}
\begin{figure}[!b]
\centering
\includegraphics[width=1\columnwidth]{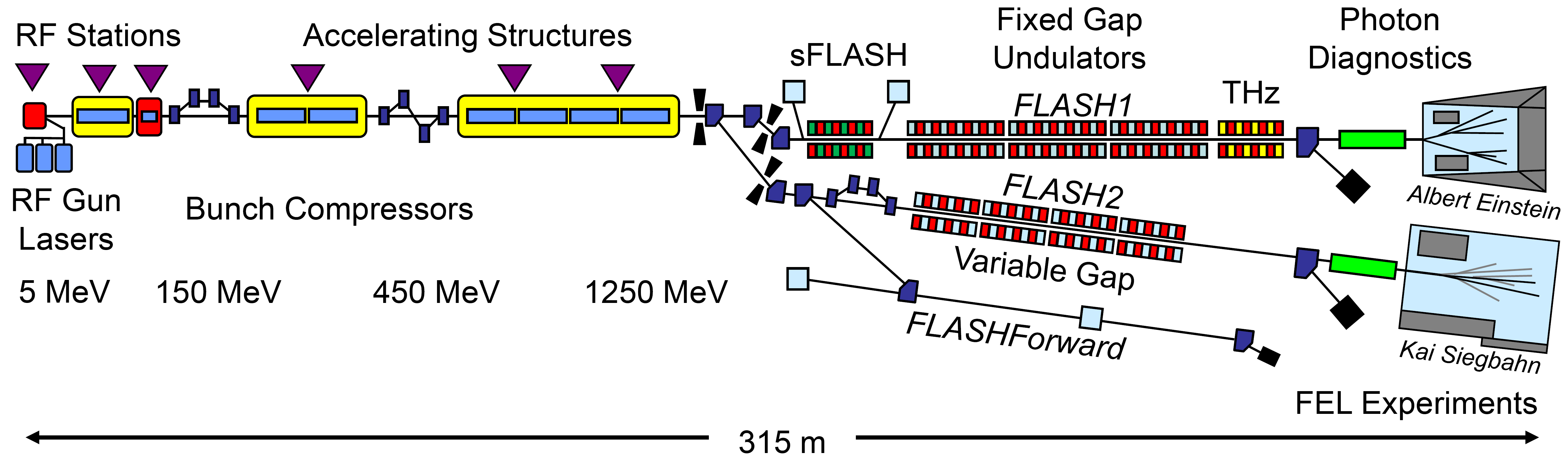}%
\caption{Schematic layout of the FLASH facility,~\cite{FLASHweb}. The electron gun is on the left, the experimental halls are on the right (components are not to scale). Behind the last accelerating module, the beam is switched between FLASH1, which is the undulator line in operation since 2005, and FLASH2 in operation since 2014.}
\label{fig:FLASHlayout}%
\end{figure}

The Free-electron LASer in Hamburg (FLASH) at Deutsches Elektronen-SYnchrotron (DESY) started user operation in 2005 as the first free-electron laser for XUV and soft X-ray radiation,~\cite{ROSSBACH20191}. It is operated in the self-amplified spontaneous emission (SASE) mode and currently covers a wavelength range from 4\,nm to 90\,nm with photon pulse durations between 10 fs and 200 fs depending on the undulator section, FLASH1 and FLASH2. FLASH1 is currently equipped with fixed gap undulator, and offers five beamlines for users in the 'Albert Einstein' experimental hall. The variable gap undulators of the 'Kai Siegbahn' experimental hall (FLASH2) enhances significantly the capacities for XUV and soft X-ray FEL users at DESY since 2014~\cite{schreiber_faatz_2015}. A schematic layout of FLASH is depicted in Fig.~\ref{fig:FLASHlayout}.

The European X-ray Free Electron Laser (EuXFEL), in user operation since September 2017, is the largest SRF accelerator and offers currently the world's highest peak brilliance in the soft and hard X-rays radiations~\cite{ALTARELLI20112845,Decking2020}. 
These two unique accelerators are based on TESLA superconducting radio frequency (SRF) cavities, housed into 8-cavity cryomodules, the building brick of these accelerators. FLASH has 7 such 1.3\,GHz cryomodules for a final energy of 1.25\,GeV, while EuXFEL has 97 for a final energy of 17.5\,GeV. 

\begin{figure}[!bt]
\centering
\includegraphics[width=1\columnwidth]{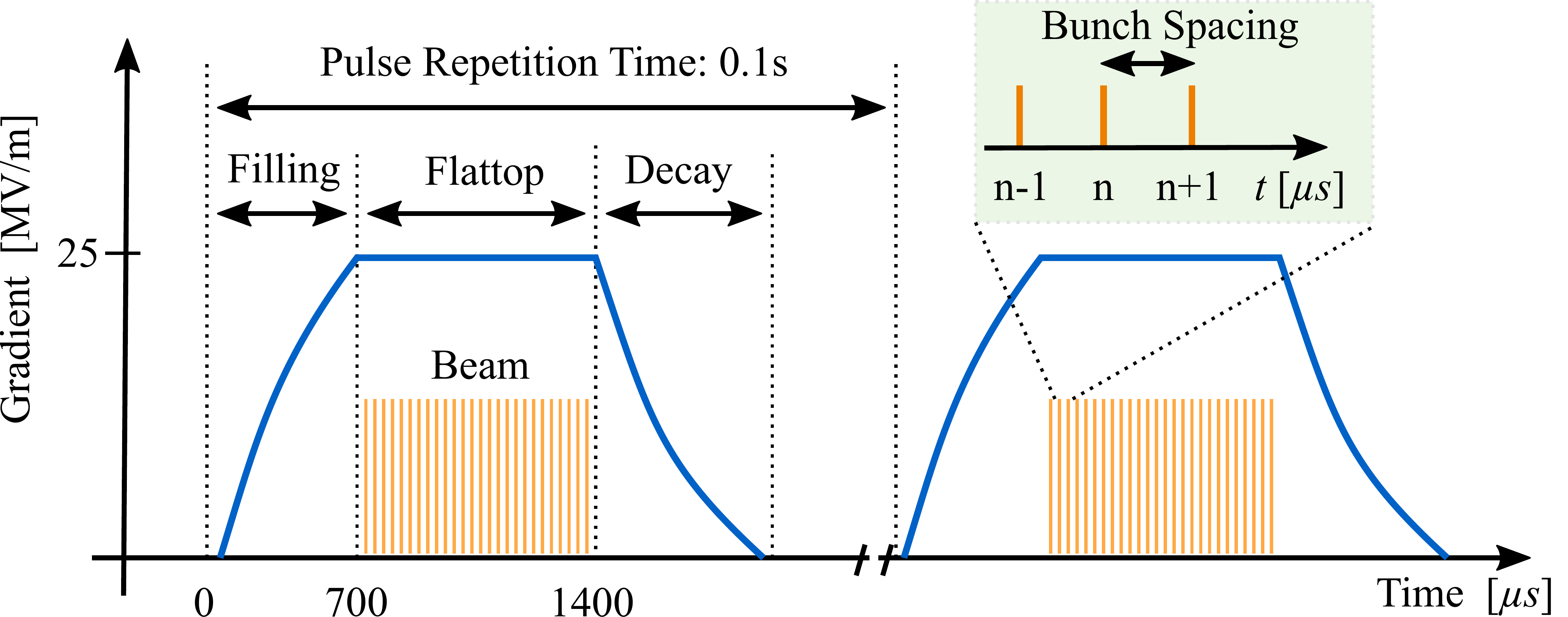}
\caption{The RF is sent to the cavity during the fill time, while the electron bunches are accelerated during the next pulse section where the field should be kept constant to provide the same acceleration to all bunches. This section is hence referred to as flat top. The following section is the RF decay.}%
\label{fig:pulsedRF}%
\end{figure}

Both FLASH and EuXFEL accelerators operate in burst mode where the SRF cavities are filled at a 10-Hz repetition rate, for a duration of 1.4\,msec per RF pulse, see Fig.~\ref{fig:pulsedRF}. The electron bunches are accelerated over these short energy bursts (ranging from $600\,\mu s$ to $800\,\mu s$), with a 1-4.5\,MHz bunch repetition rate depending on user requirements and machine capability. This pulsed mode of operation allows for very high gradients inside the SRF resonators (20-30 MV/m) while keeping the dynamic heat load within the capacity of the cryogenic plant and the liquid helium flow within the cryomodule 2-phase pipe capability. While some user experiments are only possible at the extreme high brilliance FLASH and EuXFEL can offer, other experiments would benefit from a more relaxed bunch repetition rate (kHz versus MHz) but provided continuously instead of being bunched. This calls for a different mode of RF operation, namely continuous wave (CW). The feasibility of a CW upgrade of EuXFEL is currently considered and several R\&D projects are being carried out at DESY,~\cite{Brinkmann_2014,Shu:2019ebn,Bellandi:2018gez}. Since its early days as a demonstrator of SRF accelerator-based FEL, FLASH continues to offer the opportunity to develop new techniques for SRF accelerators, which can be applied at FLASH and at XFEL. 
This contribution presents such an example of R\&D originally presented in 2016~\cite{Elmar}. In a synchrotron, the bunch-to-bunch energy spread of the synchrotron frequency can be increased by pure amplitude modulation~\cite{PhysRevSTAB.8.102801} or with a detuned cavity~\cite{Naumann:1998tw}. Pure amplitude modulation requires a broadband cavity, while operating a detuned cavity requires sufficient motor tuning margin without changing the infrastructure. The latter were tested at FLASH with its two beamlines in mid-2018 as a proof of concept for possible use in a CW EuXFEL. Such an approach have been taken into account for pure amplitude modulation maintaining the beam energy~\cite{PhysRevAccelBeams.22.110702}.

Both facilities share a common RF architecture, where a single RF source (a multi beam klystron) is used to generate a pulsed high power RF wave, split and distributed via waveguides into several cavities (up to 16 at FLASH and up to 32 at EuXFEL). The RF regulation scheme has been developed accordingly, introducing the so-called vector-sum (VS) control where the RF fields measured inside all cavities of an RF station are added up into a single complex vector used as input to the feedback controller, details of RF field sampling, processing and regulation are given in \cite{Pfeiffer:205450}. The controlled output is amplified at the single klystron and passively distributed to the multiple cavities. This RF architecture choice (driven by cost savings and technology availability at the design time) has an impact on the CW upgrade scenario. The waveguide distribution associated with each cryomodule and the concept of a single source providing power for multiple cavities are to be preserved. 

In pulsed mode, the RF energy can be modulated from pulse to pulse,~\cite{PhysRevAccelBeams.19.020703}, or within a single RF pulse, by acting on the accelerating gradient in amplitude and phase during the beam acceleration phase of the RF pulse,~\cite{Ayvazyan:ICALEPCS2015-MOC3O07}. The latter is routinely done at FLASH and XFEL where operators define several beam regions (typically 1-3) where different amplitude and phase parameters are set, providing different compression parameters for bunches as a function of their position within a single bunch train. The bunches of different beam regions can be adjusted by RF settings to have e.g. a different chirp after the bunch compression section. 
This concept allows operators to distribute subsections of a given bunch train into different beam lines with various beam properties. While this concept is well suited for pulsed-RF operation, it cannot be applied in CW. 
As an alternative, it was proposed to use some isolated cavities to provide not only energy but also amplitude and phase modulation at the level of individual bunches~\cite{Elmar}. 
This concept is explained and experimentally demonstrated in this contribution.
\\
The paper is organized as follows: 
The next section shortly introduces the digital LLRF system with multi beamline support. Section~3 discusses the operation of a detuned cavity with the normal LLRF operation scheme. Adding and tuning beam as experiment is shown in Sec.~4 with measurements at FLASH for various conditions. The last section concludes the paper and gives a short outlook to a possible CW upgrade of EuXFEL. 

\section{LLRF system and multi-beam operation}
\begin{figure}[!b]
\centering
\includegraphics[width=0.95\columnwidth]{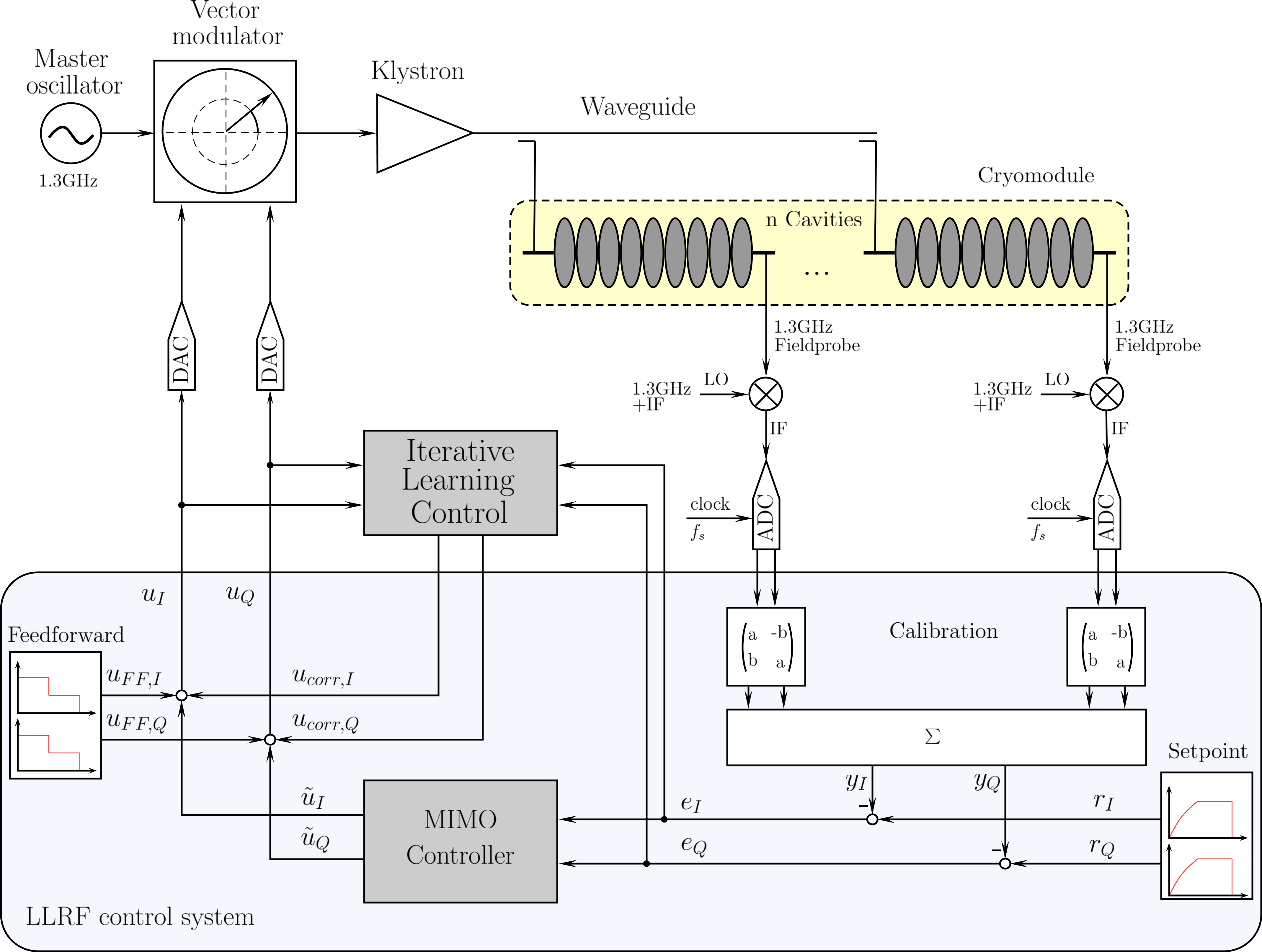}%
\caption{LLRF system overview.}%
\label{fig:LLRF_BSB}%
\end{figure}

The digital LLRF system, implemented on an FPGA based on MicroTCA.4 standard, is used to maintain and regulate the operating point of an RF station on microsecond level,~\cite{Schmidt.2014}. The LLRF regulation is depicted in Fig.~\ref{fig:LLRF_BSB}. 
The baseband drive signal in I and Q (in-phase and quadrature) is up-converted to operating frequency by a vector modulator before it is amplified from Watt to Megawatt level by a klystron providing RF to up to 32 SRF cavities at a time. The generated and stored RF field in a cavity is coupled out with a probe pickup antenna and used for signal detection. This signal is down-converted first to an intermediate frequency (IF) and finally into baseband. The sum of all probe signals is compared to a set-point and its error is processed with digital filters and a feedback controller before being added this to the nominal drive signal. This closes the feedback loop. To relax the feedback controller, the error of each pulse is used to optimize the digital drive signal from pulse to pulse to minimize repetitive errors. 
Several beam regions can be defined by splitting the digital drive signal into parts to deliver beam to more than one experiment at FLASH. A kicker-septum system has been installed behind the FLASH accelerator to split the electron bunch train between two undulator lines. The typical rise time for the kicker is in the order of 50\,$\mu s$,~\cite{Obier.2014}. This defines the lower bound on the transition time between the two beam regions. 
This transition time is also used independently of beam regions, to change RF related beam properties such as compression or energy.~\cite{Faatz:301853}. The amplitude and phase change rate between two beam regions is also limited by the small bandwidth of the SRF cavities,~\cite{Schmidt.2014}. The operation at FLASH and EuXFEL allows to have different flattops for different SASE regions. By this an independent beam tuning of e.g. FLASH1 and FLASH2 can be done.

\section{Frequency shifted cavity}
Modulating the accelerating amplitude and phase of a single bunch on a microsecond time scale can be done with different methods. Installing and operating a normal conducting cavity with a high bandwidth in standing or traveling wave mode configuration is one option. Another option that doesn't require changing the infrastructure is to operate one or more SRF cavities at a different operating frequency. 
A cavity detuned by a subharmonic frequency of the laser repetition rate will provide energy modulation to the fraction of the bunches corresponding to the subharmonic order. In the case of FLASH a 250\,kHz cavity detuning for a laser repetition rate of 500\,kHz will modulate every other bunch.
There are several aspects to take into consideration when choosing the cavity detuning: 
the detuning should be large enough to be interesting from a user perspective (i.e. laser repetition rate). The additional contribution to the vector-sum field regulation by the detuned cavity operation should be minimized (i.e., several 100\,kHz) and it should be within the tuning range of the motorized resonator frequency tuner ($<$\,500\,kHz) for typical TESLA resonators. In the remainder of this paper, we will refer to the frequency-shifted cavity as the ``slip'' cavity.
The name of ``slip'' cavity comes from the fact that the field inside the detuned cavity rotates at a frequency slower than the other cavities in a phasor diagram representation, hence providing frequency slippage. 

\begin{figure}[!b]
\centering
\includegraphics[width=0.95\columnwidth]{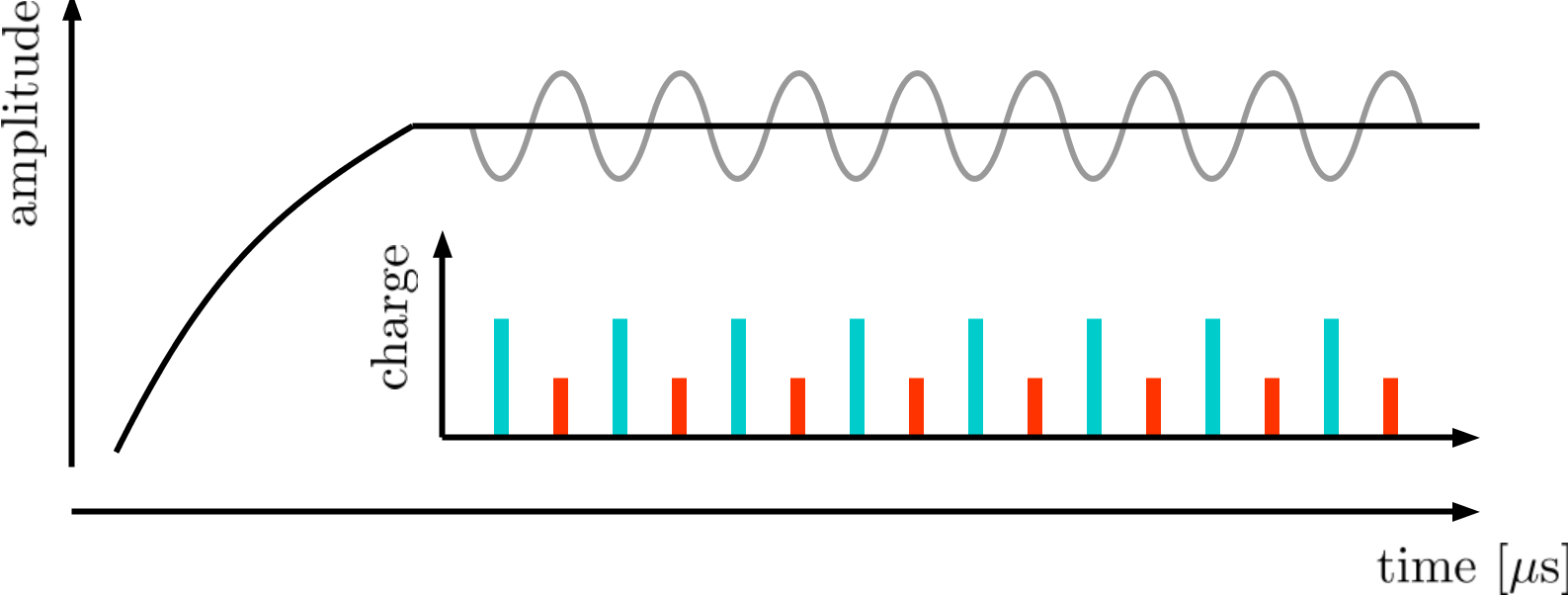}%
\caption{Amplitude (black) of the 1.3\,GHz resonance frequency and the superposed contribution of a detuned frequency (grey), both mapped into baseband representation. The bunches with repetition rate twice the detuned cavity frequency are shown in blue and red.}%
\label{fig:principle_1}%
\end{figure}

\begin{figure}[!t]
\centering
\includegraphics[width=0.77\columnwidth]{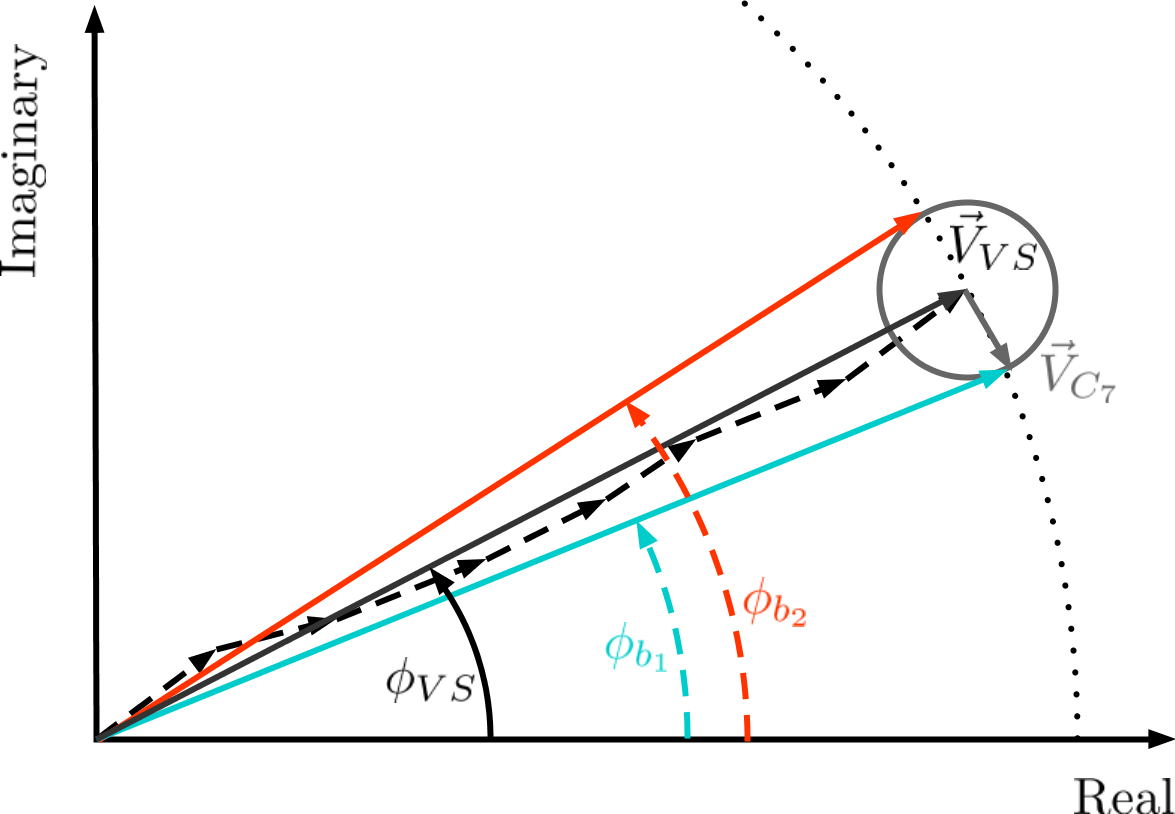}%
\caption{RF field for cavities operated in vector-sum and slip cavity (C7) in complex plane. The slip cavity is changing phase by $\pi$ for every other bunch, depicted as $\phi$ seen by first bunch $b_1$ and second bunch $b_2$ or for the given slip cavity operating frequency for every odd and even bunch, respectively.}%
\label{fig:principle_VS_1}%
\end{figure}

\paragraph{Slip cavity basics}
Bunch by bunch SASE fine tuning using a slip cavity is illustrated in Fig.~\ref{fig:principle_1}.
\\
Manipulating every bunch by the operation of such a slip cavity changes the RF property of every bunch in amplitude and phase. If for instance the second part of the bunch train (visualized by red charge components) is to be optimized, the first part (illustrated by blue bunches) will change the RF field as well. To avoid this, the second part is to be corrected by the remaining (in this case 7) cavities which are still operated in vector-sum configuration, shown in Fig.~\ref{fig:principle_VS_1}. 

\paragraph{Slip cavity operation with VS}
Consider a slip cavity operating with some amplitude and a zero phase shift between the slip cavity RF field and the vector-sum at the time of the first bunches. Every odd bunch is accelerated and every even bunch decelerated due to the energy modulation of the slip cavity. 
\\
If the slip cavity is operated with a 90\,degree phase difference from the operation of the VS cavities, a change in slip cavity amplitude would lead to a phase change for every bunch while passing all 8 cavities,~see Fig.~\ref{fig:principle_VS_1}. E.g. every odd bunch sees a decrease in phase (blue arrow) and every even bunch (red arrow) sees an increase in phase. The previous nominal operating point is depicted by the black arrow as sum of 7 cavities, each depicted as dashed arrow. Hereby, part of the bunches increase and the other part decrease their compression. To correct for this, the RF field of the blue arrow is to be corrected such that the blue and black arrows match again. This allows to tune only every even bunch without interfering with the odd bunches.  

\begin{figure}[!b]
\centering
\includegraphics[width=0.87\columnwidth]{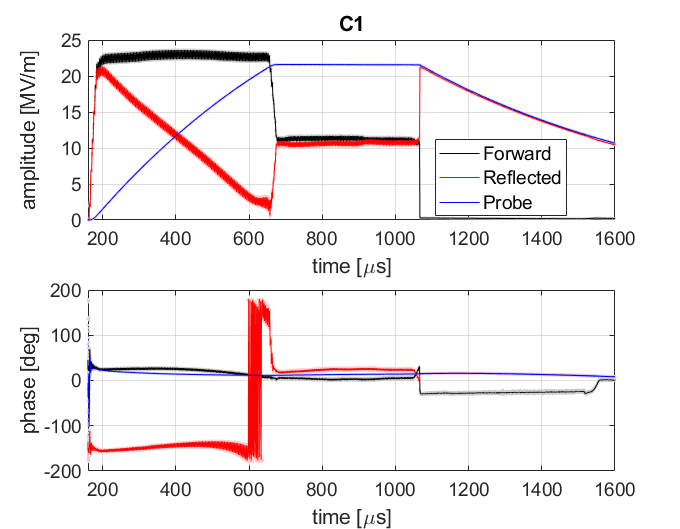}%
\caption{Cavity 1 forward, reflected and probe signal. The mean value is shown as a solid line, while the bright dots represent several consecutive RF pulses.}%
\label{fig:C1_signals}%
\end{figure}

\begin{figure}[!ht]
\centering
\includegraphics[width=0.87\columnwidth]{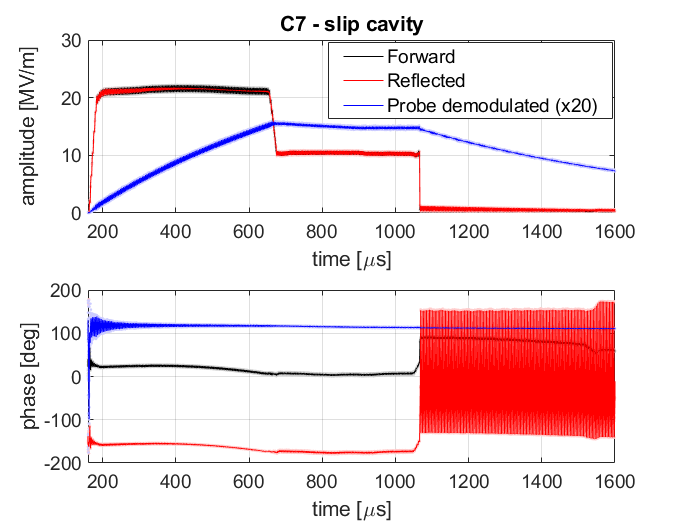}%
\caption{Cavity 7 forward, reflected and the demodulated (without -250\,kHz frequency offset) and scaled probe signal.}%
\label{fig:C7_signals}%
\end{figure}

\paragraph{Driving the slip cavity}
The LLRF system is driving 8 cavities with one klystron. Cavities 1 to 6 and cavity 8 are added to the vector-sum, while cavity 7 is operated with a frequency detuning of -250\,kHz. 
As seen in Fig.~\ref{fig:C1_signals} and~\ref{fig:C7_signals} the forward signal to all cavities is the same and differs only by its amplitude and phase from non-uniform power distribution and forward phasing. The latter can be corrected using phase shifters to compensate for waveguide length differences. 
The reflected signal differs in pulse shape since cavity~1 is filled with energy and the reflected signal reduces with external coupling factor defining the cavity bandwidth. This is different for cavity~7 operated at -250\,kHz. Here, the entire forward signal is reflected since the 1.3\,GHz baseband frequency does not match the operating frequency of the cavity. Nevertheless, the cavity is filled with RF with frequency difference of -250\,kHz, visible as decay of the oscillations on the reflected signal. Hence the cavity is filled with this frequency and the amplitude is increasing as for cavity 1. The fast phase change is related to frequency offset and given as $d\phi/dt = \Delta \omega = \omega_{0}-\omega_{0,C_7}$, with $\omega_0=2\pi \cdot 1.3$\,GHz as operating frequency and $\omega_{0,C_7}$ as operating frequency of cavity 7.

\section{Beam operation at FLASH}
\label{sec:Idea}
\begin{figure}[!b]
\centering
\includegraphics[width=1.0\columnwidth]{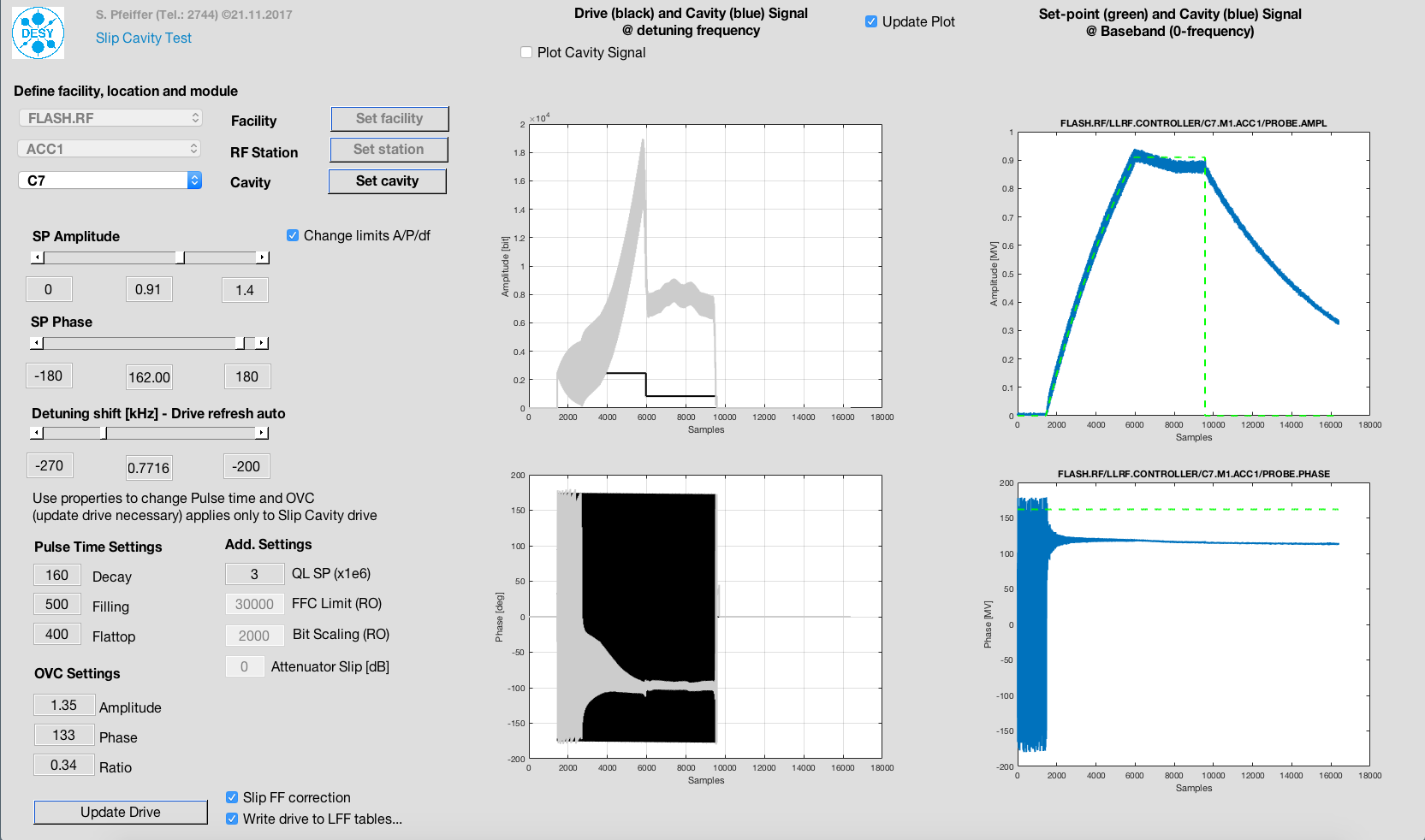}%
\caption{Graphical user interface to adjust the parameters for ACC1 - Cavity 7 during operation. The main operation parameters are amplitude and phase setpoints, detuning frequency etc. can be adjusted on left side. The middle plot shows the drive amplitude and phase for slip cavity only (black) and as modulated superposition with the ILC correction table. The demodulated slip cavity RF field probe signal is shown on the right. 
}%
\label{fig:GUI_slip}%
\end{figure}

The following conditions were used for the beam operation study at FLASH. Cavity 7 of the first accelerating structure (ACC1) was detuned by -250\,kHz and taken out of the vector-sum feedback control. Therefore its contribution is not added to the remaining 7 cavities, used for VS regulation. The previous operating amplitude for VS can also be achieved with only 7 cavities in VS. The VS regulation is operated with an iterative learning control (ILC) algorithm, reducing repetitive errors from pulse to pulse, and feedback. Learning for the ILC is switched off while  corrections are kept when operating the system with the slip cavity. This ensures that the systems does not learn from the additional drive signal. The feedback is kept active to ensure minimal error between setpoint and VS during operation. The slip cavity can be controlled in amplitude and phase with feedforward only, see Fig.~\ref{fig:GUI_slip}.  
The 250\,kHz modulated drive signal is added to the VS drive signal in baseband before up-conversion. 
To add the drive contribution of the slip cavity, we make use of the ILC table. Learning is enabled first without the slip cavity; the learned correction is then frozen and the slip cavity drive signal is added in I and Q to the total ILC correction table. ILC learning is disabled while driving the slip cavity to avoid overwriting the tables. 

\begin{figure}[!t]
\centering
\includegraphics[width=0.94\columnwidth]{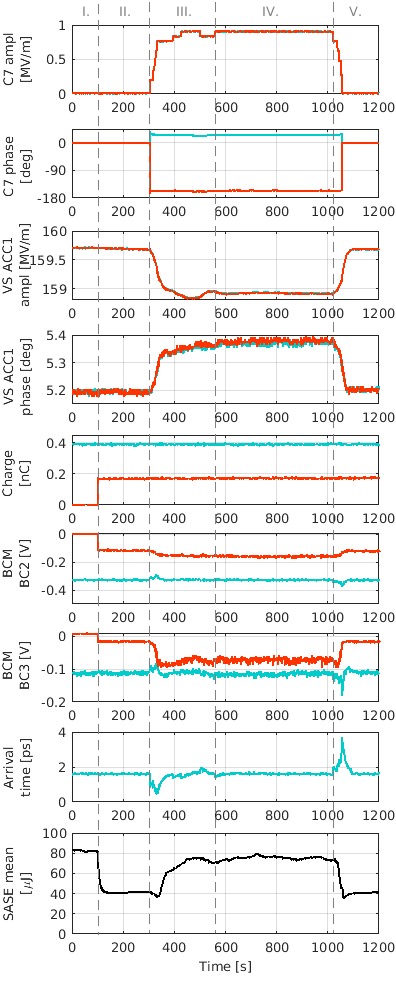}%
\caption{Main result for operation in the different time periods given in seconds with 2 different lasers and 2 different bunch charges. RF and beam signals starting from 25-Sep-2018 01:35:29. The colors corresponds to first, respectively second bunch.}%
\label{fig:Long_term_RF_beam_signals}%
\end{figure}

The time period with the operational conditions during the study time is summarized in Tab.~\ref{tab:op_cond}. Figure~\ref{fig:Long_term_RF_beam_signals} shows the main result of slip cavity operation for about 20 minutes under different conditions. 
\begin{table}[!ht]
\centering
\caption{Operational conditions}
\label{tab:op_cond}
\begin{tabular}{l|l|c|c|c|c|l}
\# & Time 	          & Las. 1 	& Las. 2 	& Slip  & SASE 	& comment \\
   & [s]              & q=0.2nC & q=0.4nC  &  cav. & [$\mu$J] &  \\
	\hline
I.	& 0-100 			    &  - 			& 500kHz 			& OFF 			& 82 			& Reference\\ 
II.	& 100-300				  &  250kHz			& 250kHz 			& OFF 			& 40 			& Laser change \\ 
III.& 300-560			  	&  250kHz 		& 250kHz 			& ON 			  & 70 			& Tuning \\ 
IV.	& 560-1000				&  250kHz 		& 250kHz 			& ON 			 	& 75 			& Operation \\   
V.	& 1000-1200				&  250kHz 		& 250kHz 			& OFF 			& 40 			& OFF \\ 
\end{tabular}
\end{table}
\\
\paragraph*{I. Setup of beam reference}
We started with zero gradient in slip cavity and restored transmission through accelerator with 7 cavities in VS with one laser operated at 500\,kHz. This data set serves as reference for nominal beam operation at FLASH. This operation is maintained by using a slow feedback system looking on the first bunch and correcting the VS amplitude and phase with beam property changes such as arrival time and compression related to operation of the slip cavity,~\cite{Kammering}. This allows independent tuning of the second beam part.
\\
\paragraph*{II. Change to 2 laser operation}
Laser 1 is switched on after 100 s producing interleaved bunches (the even bunches). The charge for these bunches is about half the charge of Laser 2. 
Switching on Laser 1 reduces the SASE intensity from 80\,$\mu J$ to 40\,$\mu J$ since this ion-current detector shows the mean pulse energy~\cite{Tiedtke_2008}. With now twice the number of bunches, assuming the same SASE intensity for the odd bunches (bunch properties, i.e. BCM and arrival time remain same) and expected no SASE intensity for the even bunches, the shown intensity is reduced by half. This reduction by a factor of two and the argumentation are based on experience at FLASH. It could not be measured directly at bunch resolution during the experiment.
\\
\paragraph*{III. Tuning and IV. Operation with slip cavity}
The slip cavity has been switched on after 300\,s. The slow RF feedback keeps the beam properties for odd bunches constant acting on the amplitude and the phase of the first accelerating structure (ACC1). It is used to compensate the additional phase shift and energy gain induced by the slip cavity. After some adaptation steps and further manual tuning of the slip cavity, the final mean SASE level of about 75\,$\mu J$ is reached. 
Because the beam parameters of the odd bunches did not change and only the even bunches have a significantly different compression signal in BC3, we conclude that the even bunches changed from almost no lasing to a SASE level which is comparable to the odd bunches, ignoring small changes in other beam parameters.
\\
\paragraph*{V. Switching off the slip cavity}
The slip cavity is switched off after 1000\,s while the beam generated with laser 1 is still present, recovering the SASE intensity level of 40\,$\mu J$ after convergence of the slow RF feedbacks. This experiment shows the usability of a slip cavity which can be used to change beam properties on microsecond level.

\section{Conclusion/Outlook}
The contribution shows the usability of beam manipulation using a cavity operated detuned with a multiple of the bunch repetition rate for possible CW operation scheme. Different user experiments can be served and beam properties can be varied on upper kHz repetition rate. This is hardly possible with bandwidth limitation of SRF cavities and no infrastructure change. We show that the operation with only 1 LLRF system is possible, however, certain algorithms like ILC need to separate the operation in frequency domain such that any superposition of VS correction by ILC and additional frequency modulation do not start influencing each other. Hereby it would be beneficial to add a second LLRF system for detuned cavity with its own high power drive chain. This would also open the possibility for fast feedback and slow correction of the detuned cavity in amplitude and phase. 
 
\section{Acknowledgment}
The authors acknowledge support from DESY (Hamburg, Germany), a member of the Helmholtz Association HGF. The work was carried out when B. Faatz was still employed as a scientist at DESY.

\normalem

\bibliography{references}

\end{document}